\documentclass[fleqn,10pt]{SelfArx} 

\usepackage[english]{babel} 
\usepackage{amsmath}
\usepackage{moresize}
\usepackage[title]{appendix}
\usepackage[stable]{footmisc}
\usepackage{fancyvrb}
\usepackage{lipsum} 
\usepackage{listings}
\usepackage{xcolor}
\usepackage{tabularx}
\definecolor{color1}{RGB}{0,0,90} 
\definecolor{color2}{RGB}{0,20,20} 

\usepackage{hyperref} 

\hypersetup{
	hidelinks,
	colorlinks,
	breaklinks=true,
	urlcolor=color2,
	citecolor=color1,
	linkcolor=color1,
	bookmarksopen=false,
	pdftitle={Title},
	pdfauthor={Author},
}

\JournalInfo{Technical Report}
\Archive{}

\PaperTitle{\Large{Faster-Than-Native Alternatives for x86 VP2INTERSECT Instructions}} 

\Authors{Guille D\'iez-Ca\~nas \\ {\normalsize guille@berkeley.edu}}

\newcommand{\keywordname}{Keywords} 

\Abstract{\footnotesize{
We present faster-than-native alternatives for the full \texttt{\small AVX512}-\texttt{\small VP2INTERSECT} instruction subset using basic \texttt{AVX512F} instructions.
These alternatives compute only one of the output masks, which is sufficient for the typical case of computing the intersection of two sorted lists of integers, or computing the size of such an intersection. 
While the na\"ive implementation (compare the first input vector against all rotations of the second) is slower than the native instructions, 
we show that by rotating both the first and second operands at the same time 
there is a significant saving in the total number of vector rotations, resulting in the emulations being faster than the native instructions, 
for all instructions in the \texttt{\small VP2INTERSECT} subset. 
Additionally, the emulations can be easily extended 
to other types of inputs (e.g.\ packed vectors of 16-bit integers) for which native instructions are not available.
}}

\begin{document}

\maketitle 


\thispagestyle{empty} 


\section{Introduction}\label{sec:intro} 




Recent Intel processors provide instructions for the computation of the intersection of short vectors of integers, grouped under the name \texttt{\small AVX512}-\texttt{\small VP2INTERSECT}~\cite{intrinsics:guide}. 
These instructions take two packed vectors of (up to) sixteen integers as input, and return masks indicating which of the entries of one vector are also present in the other vector. 
The set intersection between the two packed vectors, or simply the size of the intersection, can be easily obtained from these masks. 

Clearly, these instructions are very useful in computing the set intersection of two long vectors of \emph{sorted} integers, which can be done with the \texttt{c++} code below:
\begin{lstlisting}[label=lst:intersection,caption=Computing the set intersection of two arrays.]
 1  int32_t * compute_intersection(
 2      const int32_t * a, const int32_t * const end_a,
 3      const int32_t * b, const int32_t * const end_b,
 4      int32_t * c)
 5  {
 6    while (a < end_a && b < end_b)
 7    {
 8      __m512i va = _mm512_loadu_si512(a);
 9      __m512i vb = _mm512_loadu_si512(b);
10  
11      __mmask16 mask_a, mask_b;
12      _mm512_2intersect_epi32(va, vb, &mask_a, &mask_b);
13    
14      _mm512_mask_compressstoreu_epi32(c, mask_a, va);
15      c += _popcnt32(mask_a);
16  
17      __m512i a15 = _mm512_set1_epi32(a[15]);
18      __m512i b15 = _mm512_set1_epi32(b[15]);
19  
20      __mmask16 advance_a = _mm512_cmple_epi32_mask(va, b15);
21      __mmask16 advance_b = _mm512_cmple_epi32_mask(vb, a15);
22      a += 32 - count_leading_zeros_32bits((uint32_t)advance_a);
23      b += 32 - count_leading_zeros_32bits((uint32_t)advance_b);
24    }
25    return c;
26  }
\end{lstlisting}
where \texttt{a,b} are the input arrays of 32-bit integers, whose intersection will be stored in array \texttt{c}.
The per-iteration update of \texttt{a,b} is designed to advance them as much as possible in a single iteration, and makes use of the assumption that both input buffers are sorted.

The size of the set intersection between buffers \texttt{a,b} is simply the difference between the return value of listing~\ref{lst:intersection}, and the initial value of \texttt{c}.
Therefore, in order to compute only the set intersection size, it suffices to comment out line $14$ 
in listing~\ref{lst:intersection}.

Importantly, note that \emph{only} the first mask (\texttt{\small mask\_a}) is used, whether we are computing a set intersection, or a set intersection size.

\begin{figure*}[ht]\centering 
	\includegraphics[width=\linewidth]{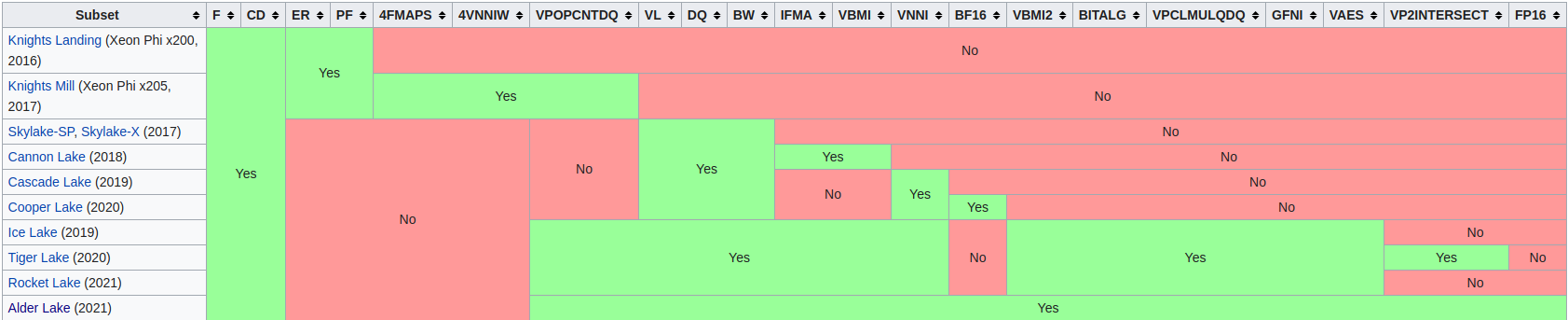}
	\caption{Support of AVX512 instruction subsets by processor architecture~\cite{wiki:avx512}.}
	\label{fig:avx512support}
\end{figure*}

Clearly, the performance of the above code is to a large degree influenced by the performance of the
\texttt{\small\_mm512}
\texttt{\small\_2intersect}
\texttt{\small\_epi32}
intrinsic, 
which is translated into the \texttt{vp2intersectd} instruction.
Note also that not all processors which have support for basic \texttt{\small AVX512F} instructions have support for \texttt{\small AVX512}-\texttt{\small VP2INTERSECT},
as seen in figure~\ref{fig:avx512support}. 
An emulation that is faster than the native \texttt{\small AVX512}-\texttt{\small VP2INTERSECT} instructions would be useful in general, and specially for those processors that don't natively support it. 
It would also potentially allow CPU manufacturers to provide similar functionality to \texttt{\small AVX512}-\texttt{\small VP2INTERSECT} without allocating silicon resources to it.

\subsection{\texttt{\small AVX512-VP2INTERSECT} instructions in practice}

Instructions in the \texttt{\small AVX512-VP2INTERSECT} subset compute intersection masks between two input vectors of packed integers (of various types and sizes), 
  as described in the reference implementation of listing~\ref{lst:intrinsic}. 
We assume that the target application of these instructions is in the computation of set intersections between two \emph{sorted} vectors of integers of arbitrary length, 
  or the computation of the size of such a set intersection, as described in section~\ref{sec:intro} and listing~\ref{lst:intersection}.
\begin{lstlisting}[label=lst:intrinsic,caption=Intrinsic for the \texttt{\small vp2intersectd} instruction operating on vectors of sixteen packed integers~\cite{intrinsics:guide}.]
void _mm512_2intersect_epi32(__m512i a, __m512i b, 
                             __mmask16* Ka, __mmask16* Kb):
MEM[Ka+15:Ka] := 0
MEM[Kb+15:Kb] := 0
FOR i := 0 TO 15
  FOR j := 0 TO 15
    match := (a.dword[i] == b.dword[j] ? 1 : 0)
    MEM[Ka+15:Ka].bit[i] |= match
    MEM[Kb+15:Kb].bit[j] |= match
  ENDFOR
ENDFOR
\end{lstlisting}

Note that operating on sorted vectors of integers has two advantages. One, it automatically eleminates duplicates. Two, and more importantly, 
  it allows the full set intersection to be computed in time \emph{linear} with the size of the input. 
If the input vectors were not sorted, we would need to check every block of sixteen integers in the first vector against every block of sixteen integers in the second vector. 
Clearly, the cost of computing set intersections on unsorted vectors of arbitrary length would be prohibitive. 

Since 
computing set intersections (or sizes of set intersections) of sorted vectors only makes use of the first output result of a \texttt{\small vp2intersectd} 
 instruction (as seen in listing~\ref{lst:intersection}), 
 we implement equivalents to all instructions in the \texttt{\small VP2INTERSECT} subset 
 that return only the first output mask. 
Note, therefore, that our implementations (except for that of section~\ref{sec:strict}) are \emph{not} drop-in replacements for the \texttt{\small VP2INTERSECT} subset. 
In fact, as shown in section~\ref{sec:strict}, the strict emulation is slower than the native instruction. 

\section{Na\"ive implementation}

Consider the \texttt{vp2intersectd} instruction within the \texttt{\small AVX512}-\texttt{\small VP2INTERSECT} instruction set, 
operating on packed vectors of sixteen 32-bit integers,
whose intrinsic's operation is 
using the notation of the Intel Intrinsics Guide~\cite{intrinsics:guide}.
In the remainder, we indistinctly refer to an intrinsic and its corresponding CPU instruction, with the understanding that all intrinsics mentioned in this paper correspond to a single CPU instruction. 

Note that for the computation of both the set intersection, and the size of a set intersection, only the first output parameter (\texttt{Ka} in listing~\ref{lst:intrinsic}) is needed, 
which is a bit-mask of the entries in \texttt{a} that are also present in \texttt{b}. 
To compute this mask, it suffices to compare \texttt{a} for equality against all entries of \texttt{b}, and bitwise-or the resulting masks. 
The ith bit of the result is set if and only if the ith entry of \texttt{a} is equal to \emph{any} of the entries in \texttt{b}:
\begin{lstlisting}[label=lst:naive,caption=Na\"ive implementation of \texttt{\small vp2intersectd}]
int16_t _mm512_2intersect_epi32_mask(__m512i a, __m512i b)
{
  __m512i b00 = _mm512_permutexvar_epi32(_mm512_set1_epi32(0), b);
  __m512i b01 = _mm512_permutexvar_epi32(_mm512_set1_epi32(1), b);
  __m512i b02 = _mm512_permutexvar_epi32(_mm512_set1_epi32(2), b);
  ...
  __m512i b15 = _mm512_permutexvar_epi32(_mm512_set1_epi32(15), b);
  
  __mmask16 m00 = _mm512_cmpeq_epi32_mask(a, b00);
  __mmask16 m01 = _mm512_cmpeq_epi32_mask(a, b01);
  __mmask16 m02 = _mm512_cmpeq_epi32_mask(a, b02);
  ...
  __mmask16 m15 = _mm512_cmpeq_epi32_mask(a, b15);
  
  return m00 | m01 | m02 | ... | m15;
}
\end{lstlisting}

The above implementation can be shown to be slower than the corresponding native \texttt{vp2intersectd} instruction.

Note that, in order to compute the output mask, it is necessary to calculate $16*16 = 256$ scalar equality-comparisons. 
Since each \texttt{\small\_mm512\_cmpeq\_epi32\_mask}
instruction performs sixteen scalar comparisons in parallel, 
there is a total of sixteen $16$-wide vector comparisons to be performed. To the best of the authors' knowledge, it is not possible to reduce this number
without introducing additional assumptions on the input vectors. 

Because the number of required comparisons is fixed, the way to speed up the above code is to reduce the number of permutations, and reduce the number of bitwise-or operations.
We note that even completely optimizing away the bitwise-or instructions (as described in section~\ref{sec:cmpneq}) is not enough to match the speed of the native \texttt{vp2intersectd} instruction.
The next section describes how to greatly reduce the number of permutation instructions, with the result that the emulations surpass the performance of the native CPU instructions.

\section{Fast emulation}

The key idea in this paper it to permute both vectors \texttt{a} and \texttt{b} of listing~\ref{lst:naive}, perform comparisons, then undo the rotations performed on \texttt{a} before combining the result
into the final intersection mask.
This reduces the number of rotation operations from 16 down to 6.

Input packed vector \texttt{a} is rotated at the 128-bit granularity only. Considering \texttt{a} as composed of four 128-bit blocks (with four 32-bit integers in each block), we compute its three possible rotations:
\begin{lstlisting}[label=lst:rota,caption=Rotation of first argument at 128-bit granularity]
__m512i a1 = _mm512_alignr_epi32(a, a, 4);
__m512i a2 = _mm512_alignr_epi32(a, a, 8);
__m512i a3 = _mm512_alignr_epi32(a, a, 12);
\end{lstlisting}
We then compute all possible rotations of \texttt{b}, at the 32-bit granularity, but only \emph{within} 128-bit blocks (note that permutations within 128-bit blocks have lower latency than those that operate across 128-bit block boundaries):
\begin{lstlisting}[label=lst:rotb,caption=Rotation of second argument at 32-bit granularity]
__m512i b1 = _mm512_shuffle_epi32(b, _MM_PERM_ADCB);
__m512i b2 = _mm512_shuffle_epi32(b, _MM_PERM_BADC);
__m512i b3 = _mm512_shuffle_epi32(b, _MM_PERM_CBAD);
\end{lstlisting}

We can now use \texttt{\small\_mm512\_cmpeq\_epi32\_mask} instructions to compute comparisons between all $16*16$ pairs of components of \texttt{a} and \texttt{b}:
\begin{lstlisting}
__mmask16 m00 = _mm512_cmpeq_epi32_mask(a, b);
__mmask16 m01 = _mm512_cmpeq_epi32_mask(a, b1);
__mmask16 m02 = _mm512_cmpeq_epi32_mask(a, b2);
__mmask16 m03 = _mm512_cmpeq_epi32_mask(a, b3);
__mmask16 m10 = _mm512_cmpeq_epi32_mask(a1, b);
__mmask16 m11 = _mm512_cmpeq_epi32_mask(a1, b1);
__mmask16 m12 = _mm512_cmpeq_epi32_mask(a1, b2);
__mmask16 m13 = _mm512_cmpeq_epi32_mask(a1, b3);
__mmask16 m20 = _mm512_cmpeq_epi32_mask(a2, b);
__mmask16 m21 = _mm512_cmpeq_epi32_mask(a2, b1);
__mmask16 m22 = _mm512_cmpeq_epi32_mask(a2, b2);
__mmask16 m23 = _mm512_cmpeq_epi32_mask(a2, b3);
__mmask16 m30 = _mm512_cmpeq_epi32_mask(a3, b);
__mmask16 m31 = _mm512_cmpeq_epi32_mask(a3, b1);
__mmask16 m32 = _mm512_cmpeq_epi32_mask(a3, b2);
__mmask16 m33 = _mm512_cmpeq_epi32_mask(a3, b3);
\end{lstlisting}

Note that masks \texttt{\small m00, m01, m02, m03} correspond to the original ordering of entries of \texttt{a}, 
while \texttt{\small m10, m11, m12, m13} correspond to an ordering of \texttt{a} that is rotated to the right by four
positions (and similarly for \texttt{\small m20, m21, m22, m23}, and for \texttt{\small m30, m31, m32, m33}, which correspond to orderings of \texttt{a} that are rotated to the right by eight and twelve positions, respectively).
Therefore the above masks cannot simply be bitwise-or'ed to obtain the final result. 

A final step needs to \emph{undo} the rotation of \texttt{a} by using $16$-bit left-rotation operations on the above groups of masks, as follows:
\begin{lstlisting}[label=lst:final:optimized,caption=Merge masks and undo rotation of first argument]
1  __mmask16 m0 = m00 | m01 | m02 | m03;
2  __mmask16 m1 = m10 | m11 | m12 | m13;
3  __mmask16 m2 = m20 | m21 | m22 | m23;
4  __mmask16 m3 = m30 | m31 | m32 | m33;
5  return m0 | __rol16(m1, 4) | __rol16(m2, 8) | __rol16(m3, 12);
\end{lstlisting}

As we will see in section~\ref{sec:cmpneq}, all but the last three bitwise-or operations (line $5$ of listing~\ref{lst:final:optimized}) can be optimized away, so we can ignore their cost.
As for the rest, we have replaced sixteen high-latency permutation operations in listing~\ref{lst:naive} by just six permutations:
three high-latency permutations in listing~\ref{lst:rota}, and three low-latency 
(\texttt{\small\_mm512}
\texttt{\small\_shuffle}
\texttt{\small\_epi32}
"within-lanes") permutations in listing~\ref{lst:rotb}, plus three bit-rotation operations and three bitwise-or operations (line $5$ of listing~\ref{lst:final:optimized}).
In our tests, this brings down the cost of a set intersection-size iteration~\footnote{Same as listing~\ref{lst:intersection} but commenting out line $14$.}
from $24.54$ clock-cycles using the native instruction, to $22.49$ clock-cycles using the emulation.
Note that the reported inverse-throughput of the \texttt{\small vp2intersectd} instruction in our test machine is $24.75$ clock cycles (as per the Aida64 benchmarking software~\cite{aida64}), 
	which means that the inverse-throughput of our test inner loop is very close to the inverse-throughput of the \texttt{\small vp2intersectd} instruction itself. 

Finally, we note that the above technique can be easily extended to different size operands (vectors of eight or four 32-bit integers), 
and to vectors of 64-bit or 16-bit integers, the last of which do not have equivalent native instructions in the \texttt{\small AVX512}-\texttt{\small VP2INTERSECT} subset.





\subsection{Final optimizations}\label{sec:cmpneq}

The intrinsic \texttt{\small\_mm512\_mask\_cmpneq\_epi32\_mask},
which codes for a \texttt{\small vpcmpd} instruction with a mask argument, 
can be seen as performing both a vector comparison, and a bitwise-and operation between the result of the vector comparison and the input mask.
Using De Morgan's laws~\cite{nievergelt2015logic}, we convert the bitwise-or operations in listing~\ref{lst:final:optimized} into bitwise-ands, which are then embedded into 
\texttt{\small\_mm512\_mask\_cmpneq\_epi32\_mask} instructions, resulting in the following final code:
\begin{lstlisting}[label=lst:final,caption=Final optimized emulation of \texttt{\small vp2intersectd} returning the first mask only.]
 1  int16_t _mm512_2intersect_epi32_mask(__m512i a, __m512i b)
 2  {
 3    __m512i a1 = _mm512_alignr_epi32(a, a, 4);
 4    __m512i a2 = _mm512_alignr_epi32(a, a, 8);
 5    __m512i a3 = _mm512_alignr_epi32(a, a, 12);
 6    __m512i b1 = _mm512_shuffle_epi32(b, _MM_PERM_ADCB);
 7    __m512i b2 = _mm512_shuffle_epi32(b, _MM_PERM_BADC);
 8    __m512i b3 = _mm512_shuffle_epi32(b, _MM_PERM_CBAD);
 9    
10    __mmask16 m00 = _mm512_cmpneq_epi32_mask(a, b);
11    __mmask16 m01 = _mm512_cmpneq_epi32_mask(a, b1);
12    __mmask16 m02 = _mm512_cmpneq_epi32_mask(a, b2);
13    __mmask16 m03 = _mm512_cmpneq_epi32_mask(a, b3);
14    
15    __mmask16 m10 = _mm512_mask_cmpneq_epi32_mask(m00, a1, b);
16    __mmask16 m11 = _mm512_mask_cmpneq_epi32_mask(m01, a1, b1);
17    __mmask16 m12 = _mm512_mask_cmpneq_epi32_mask(m02, a1, b2);
18    __mmask16 m13 = _mm512_mask_cmpneq_epi32_mask(m03, a1, b3);
19    
20    __mmask16 m20 = _mm512_mask_cmpneq_epi32_mask(m10, a2, b);
21    __mmask16 m21 = _mm512_mask_cmpneq_epi32_mask(m11, a2, b1);
22    __mmask16 m22 = _mm512_mask_cmpneq_epi32_mask(m12, a2, b2);
23    __mmask16 m23 = _mm512_mask_cmpneq_epi32_mask(m13, a2, b3);
24    
25    __mmask16 r0 = _mm512_mask_cmpneq_epi32_mask(m20, a3, b);
26    __mmask16 m1 = _mm512_mask_cmpneq_epi32_mask(m21, a3, b1);
27    __mmask16 m2 = _mm512_mask_cmpneq_epi32_mask(m22, a3, b2);
28    __mmask16 m3 = _mm512_mask_cmpneq_epi32_mask(m23, a3, b3);
29    
30    int16_t r1 = __rol16(m1, 4);
31    int16_t r2 = __rol16(m2, 8);
32    int16_t r3 = __ror16(m3, 4);
33    
34    return (int16_t)~(int16_t)(r0 & r1 & r2 & r3);
35  }
\end{lstlisting}
Note that converting bitwise-ors into bitwise-ands requires a final bitwise-not (or xor with \texttt{\small 0xffff}), as in line $34$ in listing~\ref{lst:final}.

\noindent{\bf Data dependency.} We have organized listing~\ref{lst:final} into groups of operations that can be performed in parallel with no data dependencies.
In particular, note that the sixteen comparison operations are split into groups of four instructions,
which have no data dependencies between them and therefore can be executed in parallel. 
From these groupings it should be clear that the above code does not have significant data dependencies hindering its performance.

Finally, note that the last bit-rotation (\texttt{\small\_\_rol16(m3, 12)}) is written as the equivalent \texttt{\small\_\_ror16(m3, 4)}, so as to reuse the value \texttt{\small 4} from the previous bit-rotation \texttt{\small\_\_rol16(m1, 4)}.

\subsection{Emulation with in-memory operands}

When one of the operands is in memory, for instance the second, there is a version of the naive emulation of listing~\ref{lst:naive} that loads elements from \texttt{b} and uses \texttt{\small AVX512} embedded broadcasts, as follows:
\begin{lstlisting}[label=lst:memory:naive,caption=Na\"ive emulation with second argument in memory]
int16_t _mm512_2intersect_epi32_mask(__m512i a, const int32_t * b)
{
  __mmask16 m00 = _mm512_cmpeq_epi32_mask(a, _mm512_set1_epi32(b[0]));
  __mmask16 m01 = _mm512_cmpeq_epi32_mask(a, _mm512_set1_epi32(b[1]));
  __mmask16 m02 = _mm512_cmpeq_epi32_mask(a, _mm512_set1_epi32(b[2]));
  ...
  __mmask16 m15 = _mm512_cmpeq_epi32_mask(a, _mm512_set1_epi32(b[15]));
  return m00 | m01 | m02 | ... | m15;
}
\end{lstlisting}
Note that each \texttt{\small\_mm512\_set1\_epi32} intrinsic can be folded into the corresponding comparison with an embedded broadcast. 
In fact, the \texttt{\small clang} compiler~\cite{llvm} will recognize that the permutation operations of listing~\ref{lst:naive} can be turned into implicit broadcast loads as in listing~\ref{lst:memory:naive} when the second operand is in memory. 
Listing~\ref{lst:memory:naive} is already more efficient than the corresponding native \texttt{\small vp2intersectd} instruction (note, however, that \texttt{\small vp2intersectd} only operates on registers). 

It is possible to further optimize listing~\ref{lst:memory:naive} using the idea in section~\ref{sec:cmpneq}, as follows:
\begin{lstlisting}[label=lst:memory,caption=Optimized emulation with second argument in memory]
int16_t _mm512_2intersect_epi32_mask(__m512i a, const int32_t * b)
{
  __mmask16 m00 =
    _mm512_cmpneq_epi32_mask(a, _mm512_set1_epi32(b[0]));
  __mmask16 m01 =
    _mm512_cmpneq_epi32_mask(a, _mm512_set1_epi32(b[1]));
  __mmask16 m02 =
    _mm512_cmpneq_epi32_mask(a, _mm512_set1_epi32(b[2]));
  
  __mmask16 m03 = 
    _mm512_mask_cmpneq_epi32_mask(m00, a, _mm512_set1_epi32(b[3]));
  __mmask16 m04 =
    _mm512_mask_cmpneq_epi32_mask(m01, a, _mm512_set1_epi32(b[4]));
  __mmask16 m05 =
    _mm512_mask_cmpneq_epi32_mask(m02, a, _mm512_set1_epi32(b[5]));
  __mmask16 m06 =
    _mm512_mask_cmpneq_epi32_mask(m03, a, _mm512_set1_epi32(b[6]));
  __mmask16 m07 =
    _mm512_mask_cmpneq_epi32_mask(m04, a, _mm512_set1_epi32(b[7]));
  __mmask16 m08 =
    _mm512_mask_cmpneq_epi32_mask(m05, a, _mm512_set1_epi32(b[8]));
  __mmask16 m09 =
    _mm512_mask_cmpneq_epi32_mask(m06, a, _mm512_set1_epi32(b[9]));
  __mmask16 m10 =
    _mm512_mask_cmpneq_epi32_mask(m07, a, _mm512_set1_epi32(b[10]));
  __mmask16 m11 =
    _mm512_mask_cmpneq_epi32_mask(m08, a, _mm512_set1_epi32(b[11]));
  __mmask16 m12 =
    _mm512_mask_cmpneq_epi32_mask(m09, a, _mm512_set1_epi32(b[12]));
  __mmask16 m13 =
    _mm512_mask_cmpneq_epi32_mask(m10, a, _mm512_set1_epi32(b[13]));
  __mmask16 m14 =
    _mm512_mask_cmpneq_epi32_mask(m11, a, _mm512_set1_epi32(b[14]));
  __mmask16 m15 =
    _mm512_mask_cmpneq_epi32_mask(m12, a, _mm512_set1_epi32(b[15]));

  return (int16_t)~(int16_t)(m13 & m14 & m15);
}
\end{lstlisting}
Note that in this case all comparison operations are independent, and we can therefore "chain" 
\texttt{\small\_mm512}
\texttt{\small\_mask}
\texttt{\small\_cmpneq}
\texttt{\small\_epi32}
\texttt{\small\_mask} operations in blocks of three instructions, 
as above, or in blocks of any other number of instructions. The choice to split comparison instructions into blocks of three instructions (without data dependencies within each block), 
is the one that performed best in our tests, with a set intersection-size inner loop iteration time of $22.162$ clock-cycles, 
as compared to $24.54$ cycles for the native \texttt{\small vp2intersectd} version, and only slightly beating the $22.494$ cycles for the emulation of listing~\ref{lst:final}, which does not assume any operands to be in-memory.

We finally note that \emph{which} of the two arguments is in memory is not important (either \texttt{a}, \texttt{b}, or both). If only \texttt{a} were in memory, we can swap \texttt{a} and \texttt{b} in listing~\ref{lst:memory} and use the resulting mask \texttt{Kb} to compute 
a set intersection, or a set intersection size, in a way analogous to that of listing~\ref{lst:intersection}.

\subsection{Strict emulation}\label{sec:strict}

We include here a strict emulation of the \texttt{\small vp2intersectd} instruction operating on vectors of sixteen packed 32-bit integers. 
So far we have only discussed computing the first output (\texttt{\small Ka} in listing~\ref{lst:intrinsic}) of \texttt{\small vp2intersectd}. 
The reason for this choice is that, for the common application of computing the intersection of two sets of sorted integers,
only the first output mask is needed. 
There may be applications for which computing both output masks of \texttt{\small vp2intersection} is needed. 
In this case, it is possible to strictly emulate the native instruction with some loss of performance.

\begin{table*}[t!]
  \caption{Average clock cycles spent in an iteration of the inner loop of the set intersection-size routine of listing~\ref{lst:intersection}.
}\label{tab:timings}
\centering
\begin{tabular}{|| c | c c c ||}
 \hline
 & Native (\texttt{\small VP2INTERSECT}) & Emulation & Emulation with in-memory operand \\
 \hline\hline
 \_mm512\_2intersect\_epi16\_mask & - & {\bf 31.5216} & 48.3425 \\ 
 \hline
 \_mm512\_2intersect\_epi32\_mask & 24.5413 & 22.494 & {\bf 22.162} \\ 
 \hline
 \_mm512\_2intersect\_epi64\_mask & 23.2989 & {\bf 22.2415} & 22.3273 \\
 \hline
 \_mm256\_2intersect\_epi16\_mask & - & {\bf 17.4631} & 24.6672 \\ 
 \hline
 \_mm256\_2intersect\_epi32\_mask & 16.9696 & {\bf 14.696} & 15.6312 \\ 
 \hline
 \_mm256\_2intersect\_epi64\_mask & 15.996 & {\bf 14.2548} & 14.6667 \\
 \hline
 \_mm\_2intersect\_epi16\_mask & - & {\bf 12.6286} & 14.7959 \\ 
 \hline
 \_mm\_2intersect\_epi32\_mask & 14.245 & {\bf 12.5902} & 13.9866 \\ 
 \hline
 \_mm\_2intersect\_epi64\_mask & 12.7487 & {\bf 12.5003} & 13.3062 \\
 \hline
\end{tabular}
\end{table*}

We include here the following strict emulation, 
which should be self-explanatory, except for the computation of the second output mask, 
which is simply undoing the transformations applied to the argument \texttt{b}:
\begin{lstlisting}[label=lst:strict,caption=The strict emulation of \texttt{\small vp2intersectd} is slower than the native instruction.]
void _mm512_2intersect_epi32_emulation(__m512i a, __m512i b, 
                                       int16_t * Ka, int16_t * Kb)
{
  __m512i a1 = _mm512_alignr_epi32(a, a,  4);
  __m512i a2 = _mm512_alignr_epi32(a, a,  8);
  __m512i a3 = _mm512_alignr_epi32(a, a, 12);

  __m512i b1 = _mm512_shuffle_epi32(b, _MM_PERM_ADCB);
  __m512i b2 = _mm512_shuffle_epi32(b, _MM_PERM_BADC);
  __m512i b3 = _mm512_shuffle_epi32(b, _MM_PERM_CBAD);

  int16_t m00 = _mm512_cmpeq_epi32_mask(a , b );
  int16_t m01 = _mm512_cmpeq_epi32_mask(a , b1);
  int16_t m02 = _mm512_cmpeq_epi32_mask(a , b2);
  int16_t m03 = _mm512_cmpeq_epi32_mask(a , b3);
  int16_t m10 = _mm512_cmpeq_epi32_mask(a1, b );
  int16_t m11 = _mm512_cmpeq_epi32_mask(a1, b1);
  int16_t m12 = _mm512_cmpeq_epi32_mask(a1, b2);
  int16_t m13 = _mm512_cmpeq_epi32_mask(a1, b3);
  int16_t m20 = _mm512_cmpeq_epi32_mask(a2, b );
  int16_t m21 = _mm512_cmpeq_epi32_mask(a2, b1);
  int16_t m22 = _mm512_cmpeq_epi32_mask(a2, b2);
  int16_t m23 = _mm512_cmpeq_epi32_mask(a2, b3);
  int16_t m30 = _mm512_cmpeq_epi32_mask(a3, b );
  int16_t m31 = _mm512_cmpeq_epi32_mask(a3, b1);
  int16_t m32 = _mm512_cmpeq_epi32_mask(a3, b2);
  int16_t m33 = _mm512_cmpeq_epi32_mask(a3, b3);

  *Ka = m00|m01|m02|m03 | __rol16(m10|m11|m12|m13, 4) | 
                          __rol16(m20|m21|m22|m23, 8) | 
                          __ror16(m30|m31|m32|m33, 4);

  int16_t m_0 = m00 | m10 | m20 | m30;
  int16_t m_1 = m01 | m11 | m21 | m31;
  int16_t m_2 = m02 | m12 | m22 | m32;
  int16_t m_3 = m03 | m13 | m23 | m33;  

  *Kb = m_0 | ((0x7777 & m_1) << 1) | 
              ((m_1 >> 3) & 0x1111) | 
              ((0x3333 & m_2) << 2) | 
              ((m_2 >> 2) & 0x3333) | 
              ((m_3 >> 1) & 0x7777) | 
              ((m_3 & 0x1111) << 3);
}
\end{lstlisting}
When used to compute the size of set intersections, the strict emulation above takes $31.75$ clock cycles per iteration, 
compared with $24.54$ cycles for the native instruction, and $22.49$ for the emulation that only computes the first output mask. 
Note that, when computing both output masks, we have to compute the comparison masks independently (instead of chaining them as in section~\ref{sec:cmpneq}), 
in order to combine them to compute \texttt{*Ka} and \texttt{*Kb}. Along with the final bitwise logical operations to assemble \texttt{\small *Kb} from the comparison masks, 
this accounts for the slower performance of this version.

\section{Timings}\label{sec:timings}

Timings are collected in table~\ref{tab:timings}. We measure the clock-cycle cost of an iteration of the inner loop of listing~\ref{lst:intersection} 
(with line $14$ commented out),
whose cost is very strongly dominated by the cost of the corresponding \texttt{\small VP2INTERSECT} instruction 
(for instance, we measure $24.5413$ clock cycles for the native version of \texttt{\small vp2intersectd} on sixteen packed 32-bit integer vectors, whereas the inverse-throughput reported by the Aida64 software is $24.75$).
Measurement are made on an Intel Tiger Lake i7-1165G7 (2.8GHz) CPU, which is one of the few CPUs that currently supports the \texttt{\small VP2INTERSECT} instruction subset (see figure~\ref{fig:avx512support}).

Note that in all cases the software implementation is faster than the corresponding native version. 
In only one of the cases (\texttt{\small\_mm512\_2intersect\_epi32\_mask}), the software implementation with one input argument in-memory (listing~\ref{lst:memory}) 
	is slightly faster than the implementation that operates on registers (listing~\ref{lst:final}).

Timings for the version of the software implementation that operates on packed vectors of 16-bit integers are provided for completeness, even though there are no equivalent native versions in the \texttt{\small VP2INTERSECTION} instruction subset.
Note that timings for 
\texttt{\small\_mm512\_2intersect\_epi16\_mask}, 
\texttt{\small\_mm256\_2intersect\_epi16\_mask}, and
\texttt{\small\_mm}
\texttt{\small\_2intersect}
\texttt{\small\_epi16\_mask} are more expensive than their 32-bit counterparts.
This is because 
\texttt{\small\_mm512\_2intersect\_epi16\_mask} needs to compute \emph{four times} more scalar comparison operations than 
\texttt{\small\_mm512\_2intersect\_epi32\_mask}, 
 while the relevant vector-comparison instruction 
\texttt{\small\_mm512\_cmpeq}
\texttt{\small\_epi16}
\texttt{\small\_mask}
 performs only \emph{twice} as many integer comparison operations as 
\texttt{\small\_mm512\_cmpeq\_epi32\_mask}, and therefore twice as many vector-comparison operations are needed.


%
%
%
%
%
%
%
%

\section{Conclusion}

We've shown that it is possible to implement faster-than-native versions of the instructions in the \texttt{\small AVX512-VP2INTERSECT} subset using simpler \texttt{\small AVX512F}
instructions, so long as only the first output mask is required, a case which we expect would cover most practical applications. 
Note that, in the previous sections, we suggest the use of the following name and signature for the emulated versions:
\begin{lstlisting}[label=lst:signature,caption=Signature for emulated \texttt{\small vp2intersectd} instrinsic that returns only the first intersection mask.]
int16_t _mm512_2intersect_epi32_mask(__m512i, __m512i);
\end{lstlisting}
which is significantly simpler than that of listing~\ref{lst:intrinsic}, and mirrors the naming convention of intrinsics that return a mask. 

Finally, we note that the existence of faster-than-native emulations of the \texttt{\small VP2INTERSECT} instruction subset (in the case where only the first output mask is needed), 
  suggests that the functionality of \texttt{\small AVX512-VP2INTERSECT} could be provided in software, potentially saving hardware resources.

\phantomsection




\phantomsection
\bibliographystyle{unsrt}
\bibliography{vp2inter.bib}

\begin{appendices}

\onecolumn

\section*{Compressed full source code}
\begin{Verbatim}[fontsize=\scriptsize]
#!/bin/bash
echo "QlpoOTFBWSZTWduLuSwAB5r/gHfwBABJd//vf//f67/v//9gD39vX3nvede8nBt2RXXbEqXTd3QyxVRrweTAeet61Hmdh3vNV7ZwkUAmhNNCI9Kaf
ok9RoeqeSem1EY0ACaeo9NR6E9BMJTEkaU2kU9TQPUaANDIAyDQ0DRoHqAA00HNMRkZNMmgGQ0ZDJkAAAMjTI0DCGQJT9UqlHqfqT1HqGgNAZDIAAAaMQZD
IAMgwiRSE0TQ2oaaNAANNGmgAaA0DQANAAiSECJkMp6VP9VN6o9T1GnqaHqDQaek9I0AANA9Q9R+qaAa7mQWgHiBYAF/VXufQpaShqvz0qGECum0F3JQkJU
YwyKqBTvRJqqioKwKxMkyTJMkst49ziyJBVIs8FHqSrgxCQZcoIIHHhESnlLTQLUC8uIJDOLpKC0cmcxDIiurn/LwbV/Xc8/aU/D8vRjo/cf6Tj6H4ANUKn
nh7jbXZNTYRAcGJtI8Pi/rTNjY2wOXa2MeGzz+n+AYMGGPk2hXur/UehGCh2lz4u6TY5j4RR2m3zUtoVUabW25/lXoXLnkAAAAOHOAJJJiXFyRuqGCOcFVN
0xmG5JG3JJJJJJJJJJJJJJDm+Dm+Hn7e222222MYxJJJIxjGMAIECCxxHOLbnOtccaGy97pXHca2AdPPn5p6Xef6dbr65vXPS8+n4M69hmDV6zphQ70c26d
erjCX24iV6zHisx462NPmXjmYzCJHswy1bTEObuySehsLBU1YtMAYXChUKAILObRQSSEFjeiSEFptBTTFK5AwUgkFIxSMFybJsT+2GCGIHHSDSxSAxAIsUg
EFIsUgEFIsEwk6odl1Xx0A0iPJH2IZjd78SqiOsYdbjmrtoY2NtxcH6t1Z9Xq4rEDDYULO2EcniqtNjbY3U1UmgNrH44LEw4XykB2h2h/gDmwQHwNCSyYub
ddX0HFQlKMZPdAnJYRm5+UCs68bR8Pd58bY8e5swvslxh/ES8jG+OmWcpIY0xpMISoQjqmxttSSQl91kL7beIH2/jAsPsFPzQDcm5F1B8KPaFhCB0O8pdd3
ALq6ZJC2imCSFyiJSKQaEhSFSR00kQphVAxEtSy5RogV2ZCXxTaxChcWoWoLIiWgwytG8nkuaBvA/KvyI/k3Z6TW2SeGgHNdvX22zPL0shWBUP/Cqd4hHxZ
x4pScAORzkkkjS5jhUwlLW1YZV4l8Br8hqHvacEksWaA0aA0TbsDEgjdJVpNeJDXGWtGjFgaZ4QbFjbTDaO8gYTSFol5lKktQ0kSkgp8M1d4NdLWFKhTEYL
TrHJ3MMlZJtKdV8qEa9qANrXkF5uRp5j8PSpPv+aXm65uBJ6MhQ/1RZLxniUiuPQBz+RG22+imJDbxQpHD4+7qWgQlGQIRkGQjHsZcYOoLbCgCKXdaxBiCS
SdL36t/r8ur78MMIEHDDCOEzFjYBSDQUpqq6/L1uXqZSbBsYzOt2lZcdskiu0O47ZcSbbY7Ynd2S40ixU1dK47dxJiatjcl2rpy0OR2y4khWxuDLodxDlGE
vZ+8C+4unRlZQizyePN/NmB+8/eZx+21j+3u6/n0jY2hLsPACDtPoOoa9c5vW9bly79qUsWUY2ta1qGmQzJIEILIskMmTTC1eGVXoUqRRSqiA0g/zLoJ8lU
lXHHHHKlMix08sVGlMchRIkSamru99999DTcySZ5KkmDBd3ipIOFy7uXd3XGkXblIIkkJIC6LoxRKsRZZZA0LZUohRjBQboMoXI3buYoRdNCM88QWxmGMF4
UltKYWafWr6BpBujiIiFDBjY/QTAVutjdkIiCS4iK11+CQBXMYxy2mzcTnENq3qta79d/ChTQuNkSyEhaqKMpMSA4theXWpvhsey23m+/Eo2pBsUcBEqEFg
Z5ANJAIoA6Cg7qxDNA+5rE91gHtQGCKReXKSyr6UToje5PQOOKNVtI8L83Nzc9aleBcbATJzgyc3hhhLC6FdPBKZBYdSdBQeHWEotEB6rR14GZxq+4ddZmJ
lhGsvDMQxbExayHQ3RGkFDgCC4ytDKuIBYCMEILEG1gcwHGFEwCsUEgkRAxgkBvvLZqbt9FFSjEuEodFWINl9JQBz4lwFt84EQUdPJtgWBw+4fDvguqDzz9
Tc18r1Jqz3SbRTlhEZ7SwDYA/pbc4OAB6SA5C7WuOyEN6sCuMErjIkCUGSxuMzhG8jST/9ooT9UUoZue29jsEI+FLPQ1IHeRfWRIoasRlvWS9BLMucJYQtk
lVV4UP2I078juMDDQjpw14I8isSs6reIRc7xE5CJjnvdH3FqmcMYW/LHA0KMWqiuRHLTuS2paCEZ1zzcTTSWfPWPprYQiLSoZlM1dKtzfjvmfDawYXdjRcy
rTIvkwLlnwcRLhq3AIGgB0GuPGfYAex1+Lxe74fe8vltbdhhDDE3AiXiyKVRRVFfRBwLBXzuYiUIlhufPkk89eZ7KosHo0gQgwkYx9qhGhoaGADQRCmK4Tc
wN5QEAgMGDCEJVEqG0AMauioWAkqSH9AdcICuyDBPhU9QCRDgAUYS4gwIL2oq56cq3W7u6VbbmtTcbNbrWWMYkDGJAxiQMyzkbFu7Fu7Fu7Fu7Fu7ERmYbu
4bu4a1JXg5XKAAAAMkyrJKsUysyMkkwgzMSMrHeZVdUVaK4BQwCRtttttjBjHQxJsaTGkNsaE63or5BAHQUgqYMITTs3bW7bu26blsm7w4VJiwkokosyixV
SRISMyWSYGQhAkIQJEjBYxT73wcXjr7Ozy7QeF/L4N7gxW0T1zPZBR/Bt8vHJ6go5FIdzBQgIRAgMFSMAGMAGMQIATy7f7WpnZ7LScScA7RVDIQ1Aj8eNBf
bnVVVEcWqa01EalFzYkPlC3lmDWe8GhMpnPDI1tVd48zxOcyD6VU3ahf7NbybaSz7s7TRWtjpQ7KJDXXWAE+cZVjlqDgv80dwhkePHmWg7M4z/K26dRHe00
1hToQgAtQBagJMSYJDEwGMIS0KHClmLI2AG7S63gDSFttog09s4CVawDt0OookzYkYcURLoDgrgWQYKsu74OtvLOYbnqQlOjdKImU5KLabQ7SEWjD/NJ6i2
JzB6EET0vpYPYEGIU0NKpDGU6dOmMYMY6YxgxjpjHSVJIBCyvUESIBEFCQB8XtAiFARsnxWe3ni5/PYDDpQ4EAboRBOnE69ev1QpY56aokB2r91H7X+MQNR
ewSREvRyFh7/JRwanTE4IqNjZWZvj0RtbXqxEJC4RfTy0b36Zhi6w4JsrWkIVrPpCCg0OCDo2ezlXgIRnDbcgGgRJAu+hEp5WNQrIYu8HTC296RMYlxlmhR
Zqtay1qt3bdADbSbLNWarWstaujlZu8TGQ1WZ5VmK3RAq8vBQuKLdVoAKEAuWVVkAUjjI78HHIUULcLZ5X7BVCFXQMKqlN5lONXCtAKu08Q9FdQlHhvJRrW
DKuWqzFbce2ZbcOWYZMlKSSUxkyUpJJJkyUpJJJJIBJIBJIJiSBJSBZ4dZxZyLgyrj5GXEqYSBN226MNrAEr6BKAi4lKQVQZkNxQkloGPp1pOLBcyCYGU0E
475cho8TN0UTDBVJsPeKAHdcAXwPD5M+Bg2NAASMkIvAGIJ4d3THEHzWDaic2BkeqgIJIM3hzNYfHjuBgbkhT3oXJpbNQokGhG4MjFLczQooJUJOo7sXANV
Rd9cdUFG4YIGuqCgwO6za7zWEQarvaWzJ4VrICxACt4yKm/IsBqBUANaE3gEVKR8caOcGiYN4wB79UZQcNjIs6NoNc65LeKVICkgoQ5RIkAmwukKutwzDaG
ghFef680tiaHFIHwnXqXz6KNd/geaSZjRjzKo9p3mhVUKVDeAQDzj6zpn0nNKmT2HG+730f/HkQhmhZIMiQH2f8XckU4UJDbi7ksA="|base64 -d|bzcat
\end{Verbatim}

\end{appendices}

\end{document}